# Application of the NIST AI Risk Management Framework to Surveillance Technology

**Nandhini Swaminathan**
University of California, San Diego, nswaminathan@ucsd.edu

**David Danks**
University of California, San Diego, ddanks@ucsd.edu

This study offers an in-depth analysis of the application and implications of the National Institute of Standards and Technology's AI Risk Management Framework (NIST AI RMF) within the domain of surveillance technologies, particularly facial recognition technology. Given the inherently high-risk and consequential nature of facial recognition systems, our research emphasizes the critical need for a structured approach to risk management in this sector. The paper presents a detailed case study demonstrating the utility of the NIST AI RMF in identifying and mitigating risks that might otherwise remain unnoticed in these technologies. Our primary objective is to develop a comprehensive risk management strategy that advances the practice of responsible AI utilization in feasible, scalable ways. We propose a six-step process tailored to the specific challenges of surveillance technology that aims to produce a more systematic and effective risk management practice. This process emphasizes continual assessment and improvement to facilitate companies in managing AI-related risks more robustly and ensuring ethical and responsible deployment of AI systems. Additionally, our analysis uncovers and discusses critical gaps in the current framework of the NIST AI RMF, particularly concerning its application to surveillance technologies. These insights contribute to the evolving discourse on AI governance and risk management, highlighting areas for future refinement and development in frameworks like the NIST AI RMF.

## 1 Introduction and Background

Surveillance technologies are increasingly widespread in both public and private spaces, often being developed and deployed with little engagement from relevant stakeholders. Most notably, the individuals subject to the surveillance technology are rarely included in creating that technology. As an illustration of both prominence and controversy, one may consider the AI system developed by Clearview AI Inc. to monitor and record the activities of individuals and groups, including rapid face identification. Their system has come under close scrutiny for the ways that the organization scraped images and training data from the Internet; the company is currently under investigation in multiple jurisdictions for scraping billions of images from social media sites without users' consent [1, 2], and other companies like Facebook, Twitter, Venmo, and Google have issued cease and desist letters citing violations of their terms of service [3]. At the same time, the overall design of their system is increasingly common in surveillance systems. Utilizing advanced technologies like facial recognition and GPS tracking trained using very large datasets (e.g., potentially over 40 billion facial images [4]; see Figure 1), these systems manage sensitive data, including personal images and biometrics, whether obtained from public or private sources [5, 6]. Clearview and other surveillance companies have argued that these data play a vital role in supporting law enforcement, government, and military operations, particularly in crime investigations and public safety enhancement [7, 8]. Similar claims have been made by employers using face recognition to monitor employees, apartment managers using surveillance AI systems for their properties, and so forth. However, there are equally many clear risks, ranging from privacy concerns when there are security breaches [9] to discriminatory or illicit misuse of the systems [10, 11].

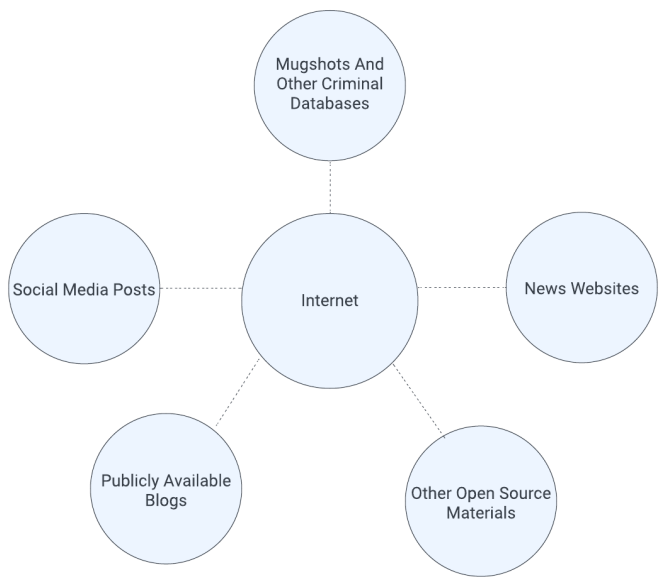

**Figure 1. Overview of Data Sources for Clearview AI's Facial Recognition Software as Described in Official Documentation**

    In the broader context of AI surveillance technologies, Clearview AI is just one of many players, leading to a pressing need to address the potential harms of such technologies. A key tool in this effort is the National Institute of Standards and Technology's AI Risk Management Framework (NIST AI RMF). This framework, while voluntary, provides essential guidance on AI risk assessment and management for AI system developers, deployers, users, and evaluators. It is particularly relevant in high-stakes scenarios, as the focus on the full lifecycle of AI system development and deployment helps to ensure that a range of possible risks and responses are considered. The framework can be used to identify, assess, and mitigate risks inherent in AI systems, including both critical potential harms such as privacy violations, ethical risks, and threats to civil liberties and also potential benefits. That is, it is important to note that the word 'risk' encompasses (for the AI RMF) both positive and negative potential outcomes.

    The NIST AI RMF is structured around four core functions: GOVERN, MAP, MEASURE, and MANAGE (see Figure 2) [12]. Each function is specified at a relatively high schematic level; the details must be determined for the particular (proposed) uses of AI technology in specific contexts. The GOVERN function establishes policies, procedures, and practices in alignment with the organization's guiding principles and strategic objectives. The MAP function involves identifying and analyzing AI-related risks and their potential ramifications. The MEASURE function brings rigor to risk management by employing various quantitative and qualitative tools to assess the actual risks, impacts, and tradeoffs. The MANAGE function synthesizes insights from the preceding functions to prioritize and address AI risks effectively.

    Although each application of the AI RMF must be tailored to the specific use-technology-context case, there are typically similarities across different analyses in the same sector (captured in what NIST calls 'RMF profiles'). We thus

develop an analysis for a specific type of facial recognition system—specifically, Clearview AI's system, as many details have been publicly disclosed—to identify risks and responses (both positive and negative) for surveillance technologies. This case study not only examines compliance with legal standards but also explores the equilibrium between technological advancement and ethical obligations to protect individual privacy. Implementation of this analysis would also raise challenges of resource allocation, technological constraints, and reconciling the interests of diverse stakeholders.

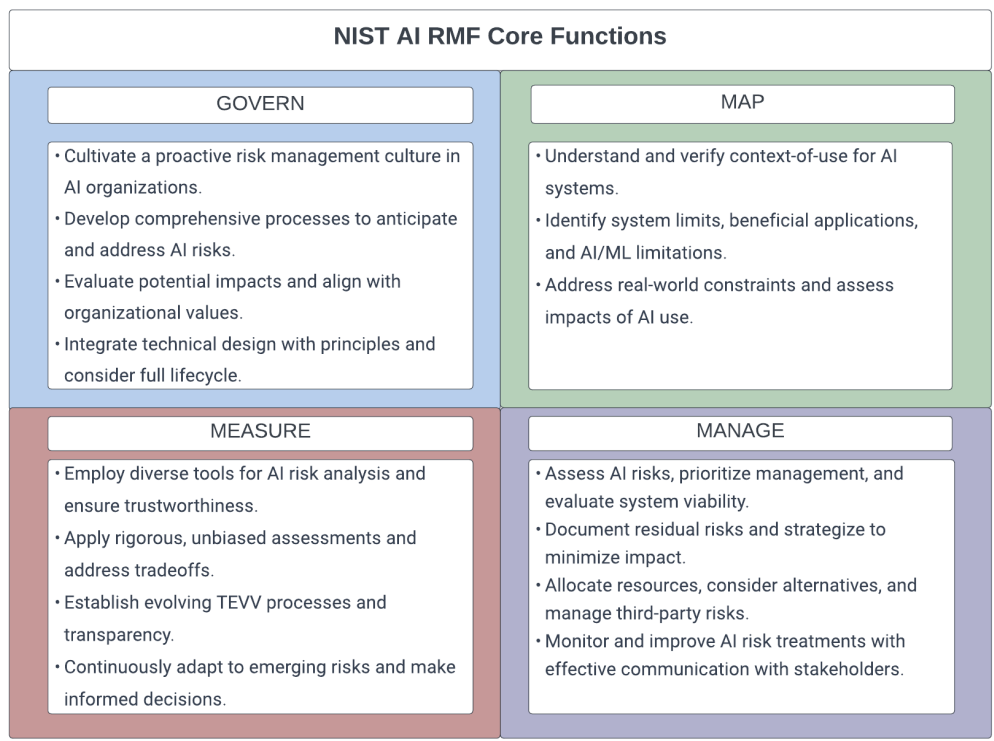

Figure 2. High-Level NIST Risk Management Framework Overview: Core Functions

## 2 Methodology: Implementing the NIST AI RMF for Surveillance Technology

### 2.1 AI System Assessment

When assessing the risks associated with AI systems used by surveillance companies, a comprehensive and structured approach is crucial. Surveillance technologies, such as those deployed by companies like Clearview AI, often process vast amounts of sensitive data, including biometric information. These systems raise significant concerns around privacy, ethics, and potential misuse, necessitating a detailed risk assessment process. A key element in this assessment is the implementation of the MAP function, a framework that provides a systematic way to evaluate and manage these risks. The first step—**Map 1 (Context is established and understood)**—requires a clear articulation of the contexts and goals of the use of the AI system. For example, Clearview AI's algorithm is designed for law enforcement and aims to match faces from a vast database of images, often controversially scraped without consent, to determine the identity (or small set of potential identities) associated with given images.

The next parts of the MAP function turn to the technology itself. In **Map 2 (Categorization of the AI system is performed)**, we must understand the AI in terms of its technical architecture and functionality. Upon analyzing the organization's patents, we observe that Clearview AI's facial recognition technology employs a sophisticated system comprising two main components: a scalable pipeline for preparing the training data and a gradient accumulation approach to deep neural network training [13, 14]. The training data pipeline gathers facial images from the internet, categorizing them into identity-based clusters. Further refinement is achieved through second-order clustering, combining clusters with high similarity scores within the same threshold range. Their patented use of gradient accumulation for neural network training where activations of initial network layers are discarded during the forward pass and recomputed in the backward pass further optimizes this training process.

In **Map 3 (AI capabilities, targeted usage, goals, and expected benefits and costs are understood)**, the focus shifts to a comprehensive understanding of the AI system's capabilities, targeted usage, goals, and the expected benefits and costs. For instance, Clearview AI's primary goal is to provide a robust tool for law enforcement agencies to identify suspects using facial recognition technology. MAP 3 also critically examines the associated risks and costs, in line with the RMF's approach that includes both benefits and potential risks in the analysis. This includes an in-depth consideration of privacy concerns, the possibility of biases in the facial recognition process, and the ethical questions raised by amassing a vast database of images without user consent. At this stage, we do not need to come to a final decision; rather, the goal is simply to understand how technical and performance benefits might tradeoff against ethical, legal, and societal considerations.

As this overview indicates, MAP analyses can quickly become complex and time intensive. To simplify the process, we propose using an "AI System Risk Categorization Matrix" (ASRCM; see Table 1), a streamlined framework inspired by the MAP 2 and MAP 3 sub-functions and the NIST SP 800-60 Vol 2 from cybersecurity risk assessments [15]. The ASRCM facilitates efficient evaluation and categorization of AI risks while upholding the principles of the MAP functions. It offers a user-friendly way for organizations to identify and prioritize risks, enhancing AI risk management. The ASRCM considers the following factors:

**Table 1: AI System Risk Categorization Matrix (ASRCM)**

| Attribute | Categories | Clearview AI's facial recognition system (2020) |
|---|---|---|
| Use case | Description | Primarily for law enforcement to identify suspects by matching faces with a vast internet-sourced database. |
| Potential impact | Health/Social/Economic | **Health**: Minimal. <br> **Social**: Significant, raises privacy and consent issues. <br> **Economic**: Impacts security sector, potential cost savings and privacy breach costs. |
| Data sources | **Sensitivity**: Sensitive/Non-sensitive <br> **Accessibility**: Restricted/Public <br> Regulatory **Compliance**: Compliant/Non-compliant | **Sensitivity**: Highly sensitive (biometric data). <br> **Accessibility**: Restricted to approved entities. <br> **Regulatory Compliance**: Contentious, varies by region. |
| Level of complexity | White/Gray/Black box | Black box - algorithmic specifics are undisclosed. |
| Regulatory requirements | Yes/No | Yes, subject to privacy and biometric data laws, compliance varies. |
| Level of autonomy | Low/Mid/High | High; Requires human oversight for result interpretation. |

We propose that the ASRCM can enable organizations to quickly identify relevant aspects of the AI system, and thereby map and prioritize risks, particularly in contexts where the AI system's impact is extensive and multifaceted. This approach can help stakeholders without deep technical expertise to better identify questions to ask about potential risks, whether harms or benefits.

In **Map 4 (Risks and benefits are mapped for all components of the AI system)**, we need to understand how the potential harms and benefits arise from different parts of the broader sociotechnical system that contains the AI technology. For example, Clearview AI's facial recognition technology poses significant civil liberties threats, with its capability to identify individuals from online images raising concerns about surveillance and consent. This particular risk primarily arises from the possibility of human misuse of the system. In contrast, privacy concerns arise from both possible human misuse and also the existence of a very large labeled image database that could be hacked or leaked. Or consider algorithmic biases in the system, especially against women and people of color, that can lead to discriminatory outcomes, and are likely due to biases in the training data (coupled with choice of learning algorithm). Of course, the technology also offers potential benefits like enhancing law enforcement efficiency by aiding in quick identification of suspects in criminal investigations.

Finally, **Map 5 (Impacts to individuals, groups, communities, organizations, and society are characterized)** considers the real-world impacts of the risks of potential harms and benefits. Clearview AI's technology impacts individuals through potential invasions of privacy and risks of misuse, such as harassment or wrongful identification. Groups, especially marginalized communities, face the risk of discrimination due to biases in the technology. For law enforcement and organizations, while the technology offers operational efficiency, it also necessitates ethical considerations and cybersecurity vigilance. At a societal level, the widespread use of such technology calls for more stringent privacy laws and regulations to protect individual rights and manage the balance between security and personal freedoms.

By systematically evaluating the AI systems in operation according to these key factors, organizational teams can gain valuable insights into the characteristics and intricacies of their technological assets. Moreover, this comprehensive assessment facilitates identifying and managing potential risks, enabling teams to proactively address vulnerabilities and enhance their AI implementations' overall security and effectiveness.

## 2.2 Mitigation Plan

The AI RMF is not intended solely to diagnose potential issues but also to guide responses, including efforts to minimize the potential for harm and maximize the potential for benefit. An organization using the AI RMF should thus also develop a comprehensive mitigation and remediation strategy. We propose that such a plan should include the following recommended technical measures to address the company's challenges:

**Privacy and Legal Concerns:** The organization should address privacy and legal concerns by adhering to jurisdiction-specific data protection laws, providing a comprehensive privacy policy that is understandable by relevant stakeholders, and practicing data minimization. Obtaining truly informed consent (not simply clicks of "I agree") [16], offering opt-out mechanisms, and implementing robust data security measures can further safeguard user privacy. Regular audits, transparency in disclosing data sources and usage, assigning a dedicated privacy officer, and collaborating with regulatory bodies will help ensure ongoing compliance with evolving privacy norms and legal frameworks. We acknowledge that companies might resist implementing such measures due to a belief that these efforts would reduce performance or profit. The AI RMF does allow that profit can be a benefit of a system, and so these factors can be incorporated into the analysis. However, we note that companies would need to determine whether there is *actually* a performance/profit vs. ethics tradeoff, not simply assume that there must be.

**Biases and Inaccuracies:** Clearview's facial recognition system has faced criticism for biases and inaccuracies, especially when identifying people of color, women, and other marginalized groups [11]. To tackle these concerns, the organization should curate diverse training datasets and continuously evaluate system performance against fairness metrics. They can participate in benchmark tests such as the NIST Face Recognition Vendor Test (FRVT) [17] to measure their system's accuracy and bias. Additionally, it is important to report all relevant statistics (i.e., detection rate, false positive rate, etc.) to present a complete view of the system's accuracy [18]. Furthermore, as suggested by MEASURE 2, conducting algorithmic audits to identify and mitigate biases, perhaps also engaging external experts for third-party evaluations can further enhance accuracy and fairness. Clearview AI can foster trust in its technology and minimize the risk of wrongful identifications by emphasizing transparency, accountability, and adhering to industry standards.

**Data Breach:** In February 2020, Clearview AI experienced a data breach [9] exposing customer information and other sensitive data. Clearview AI should adopt encryption methods like AES [19] for data at rest and TLS 1.2 [20] or higher for data in transit to prevent future breaches. Implementing robust access controls, conducting ISO 27001-compliant security audits [21, 22], and providing employee cybersecurity training can enhance their proactive threat management. Additionally, an incident response plan aligned with the NIST Cybersecurity Framework [23] will bolster their defenses against breaches.

**Unethical Business Practices**: To mitigate potential misuse, Clearview AI should choose a set of strict usage policies that align with reputable regulations or guidelines. One set of possibilities would come from legal frameworks, such as the EU Artificial Intelligence Act [24] or international AI guidelines such as the OECD Principles [25] or Universal Guidelines for AI. Alternately, Clearview (or other surveillance technology company) could articulate its own set of clear principles and practices that will guide their acceptable use policies. The key is for Clearview AI to select a framework that suits its operational context and goals, and then broadly publicize and adhere to it. By enforcing compliance through regular audits, implementing data access controls, and engaging with regulators, Clearview AI can ensure transparency and adherence to regulations. Amid a patchwork of U.S. state laws and federal proposals, including the Facial Recognition and Biometric Technology Moratorium Act [26], Clearview AI should proactively comply with state-level biometrics laws, which mandate consent and detailed privacy notices, and maintain robust privacy and data security measures that align with existing and emerging regulations in the evolving legal landscape.

## 2.3 Adaptation and Implementation Strategy

The previous section laid out a long list of issues and changes that are suggested for Clearview AI by the application of the AI RMF. Importantly (and in contrast with the EU AI Act), the conclusion is *not* that Clearview AI should simply shut down and not offer its technology. On our analysis, there is a potential path forward towards ethical development and use of this technology, though it has many hurdles. Given the complexity of the challenge, and in light of the likely temptation to cut corners, we suggest that a company in this position should engage outside, third-party consultants (or other independent agents) [27] to expedite adaptation to the guidance of the framework. This move would help establish credibility and ensure impartiality while still ensuring that the changes are sensitive to the organization's AI systems, data privacy requirements, and specific risk factors (unlike, for example, an arms-length auditor or certifier).

Working alongside the consultants, a surveillance technology company will be able to define its risk tolerance, prioritize AI risk management activities (GOVERN 1.3), and establish processes for periodic review and monitoring of risk management (GOVERN 1.5). This process is supported by documentation of the intended purpose, potential benefits, and risks of the facial recognition system (MAP 1.1, 1.5). Technical techniques like Failure Modes and Effects Analysis (FMEA) [28] could also help to develop an increased employee and user understanding of the key risks and corresponding mitigations. The organization could also collaborate with interdisciplinary AI actors and domain experts to assess AI risks and understand the context of deployment (MAP 1.2, 1.6) and elicit system requirements while assessing privacy, fairness, and bias-related risks (MAP 1.6, 5.1).

A different set of challenges arise in the development and implementation of the MEASURE functions. This company needs to be able to evaluate AI system performance and risks, which will require the selection of appropriate metrics and assessment tools to measure AI risks and system performance, including accuracy, fairness, and privacy (MEASURE 1.1). This step may also require the development of novel measures for real-world assessment, particularly since some evaluations should be conducted in conditions similar to deployment settings (MEASURE 2.3). The organization must evaluate the system for safety risks, fairness, and bias, documenting residual risks and limitations (MEASURE 2.6, 2.11). And measurement continues even after deployment, as monitoring plans must be designed and implemented to ensure that there are vehicles to obtain user input, incident response, and response patterns (MEASURE 4.1, 4.3).

And, of course, a key goal of these changes is to ensure that the company can satisfy the requirements of the MANAGE functions. In particular, the organization must implement robust data privacy and security measures alongside other responses to the high-priority risks of harm (MANAGE 1.2, 1.3, 2.1). Similarly, they must implement procedures that sustain deployed AI systems and ensure that system performance aligns with intended use (MANAGE 2.2, 2.4), including responses to, and recovers from, previously unknown risks (MANAGE 2.3). Depending on the structure of the effort, the organization may also need to manage risks associated with third-party resources, including pre-trained models and third-party AI technologies (MANAGE 3.1, 3.2). Feedback and measurable performance improvements are integrated into AI system updates (MANAGE 4.2, 4.3) to ensure constant and continuous improvement.

### 2.3.1 Ethical Data Practices

In thinking about adaptations and mitigations, there are two areas that are particularly salient for surveillance technology companies: data practices and assessment processes (Section 2.3.2). On the first point, Clearview AI has faced significant criticism and legal challenges regarding its data collection and usage policies. Its practices have been deemed a clear violation of privacy by data protection authorities in various countries, leading to significant fines and orders to cease data collection and delete existing databases. To handle this, it must implement the following steps:

- Data Collection and Quality Assurance:
    - Embed written procedures into the AI data collection process, ensuring consistency and minimizing errors.
    - Use qualified personnel for data collection, maintaining oversight to prevent unauthorized changes and ensuring data integrity.
- Governance and Policy Development:
    - Develop transparent policies that comply with legal and regulatory standards, incorporating data quality requirements directly into the governance framework (GOVERN 1.1, 1.2). Furthermore, incorporate data privacy and quality checks into the CI/CD (Continuous Integration/Continuous Deployment) pipeline to ensure new code adheres to these standards.
    - Organize training sessions for all team members on data collection standards and ethical considerations. Implement a certification program to ensure all operators are well-versed in these standards.
- Risk Management and Data Safeguards:
    - Prioritize AI risk management activities that include data quality checks and the safeguarding of sensitive data against unauthorized changes (GOVERN 1.3, 1.5). Tools like SonarQube for code quality checks and DAST (Dynamic Application Security Testing) for security vulnerabilities[29] can be used.
    - Schedule periodic audits of both the data and the systems used for data processing. Use automated tools for compliance checks and perform manual reviews regularly.
- Secondary Data Utilization:
    - When using data collected by others, understand the methodology by examining the methods used to collect the external data. This involves understanding the data sources, the collection process (like surveys, web scraping, etc.), and any preprocessing steps applied. Additionally, identify potential biases or limitations in the data.
    - Establish automated checks to verify that external data meets the organization's quality and ethical standards. Implement a feedback loop from these checks to continuously improve data integration processes (MANAGE 3.1, 3.2).
- Stakeholder Engagement and Feedback:
    - Incorporate feedback mechanisms for both primary and secondary data use, allowing for adjustments based on stakeholder insights.
    - Engage with external data sources to potentially assist in enhancing their data quality and management systems (MANAGE 4.2, 4.3).

### 2.3.2 Assessment Process for Ethical and Responsible Use of Facial Recognition Technology

A significant task for the company would be the establishment of governance principles and policies that comply with legal and regulatory requirements such as data privacy and ethical standards (GOVERN 1.1, 1.2). However, the establishment of principles is insufficient, as they must also be followed in practice. As noted above, we propose that third-party auditors or consultants would be better agents for assessments (compared to internal teams within the company), but given the variability of regulations across regions, the consultants would need to tailor assessment of the policies to align with the specific legal nuances of each jurisdiction. Additionally, to mitigate misuse, consultants would need to advocate for the role of a certified operator for each client, where the operator, well-versed in legal and risk aspects, can oversee the system's use. There are thus many steps in assessment of the company's performance in terms of maintaining its ethical principles, starting from the initial assessment of the importance of purpose and progressing through various crucial checkpoints such as accuracy, potential for error, privacy concerns, and practicality, leading to a final decision on whether to use the facial recognition technology. Following these steps will ensure the technology is used responsibly and ethically:

- Evaluate the ethicality and legality of the client's purpose in using FR. This determines whether the use justifies the technology's deployment.
- Respect privacy norms and consent mechanisms, especially in data enrollment and sourcing, ensuring there is no unwarranted intrusion into personal spaces.
- Define the scope of searches, whether they are targeted or general, and the level of suspicion required for individual searches.
- Evaluate image quality and accuracy and check for potential error and bias
- Assess potential for abuse/misuse by client
- Scrutinize the data sources, whether governmental or private and ensure they are used for legitimate purposes without infringing on protected activities.
- Assess if other identification means are overly burdensome or if FR is the most practical option in time-sensitive situations.
- Establish clear policies for how data collected through FR is shared, with whom, and for what purposes, maintaining transparency in its usage.
- Ensure compliance with all relevant laws and regulations and adhere to ethical standards and best practices in the deployment of FR technology.

Furthermore, the organization could also utilize the Video Privacy Protection Act (VPPA).119; the act requires a warrant before a video service provider may disclose personally identifiable information to law enforcement. Implementing this would ensure that the client obtains a legal warrant prior to approaching the company, minimizing the potential for misuse.

## 2.4 Workshops and Stakeholder Engagement

One of the most interesting aspects of the AI RMF is the emphasis on stakeholder engagements. Multiple steps in the AI RMF emphasize the need to connect with a range of stakeholders (not just customers and developers) to better understand the potential impacts of the AI technology. In particular, multiple GOVERN functions (1.5, 3.1, 5.1, 5.2) can be naturally understood to offer valuable guidance on effective stakeholder engagement, from outreach to user testing to post-deployment monitoring. We focus here on workshops that should be organized for leadership and stakeholders to co-develop a thorough understanding of the surveillance AI system set for deployment, perhaps using the lens of the NIST AI RMF. Surveillance technologies frequently involve stakeholders with opposed values (e.g., privacy vs. security), and venues such as workshops provide the opportunity for deliberation and discussion about the tradeoffs and possible paths forward. These workshops could prove effective by actively engaging stakeholders in meaningful dialogue about the AI system, significantly deepening their understanding of its operation. The sessions should develop a detailed implementation plan, concentrating on technical requirements, legal compliance, and ethical considerations. Additionally, to further promote and enhance transparency, we recommend the release of specific documents to stakeholders, as outlined in Table 2.

**Table 2: Documents for Stakeholder Engagement**

| Stakeholders | Documents |
|---|---|
| End users and Data subjects | - Clearview's AI system Privacy Policy <br> - Explanation of how the system works, its purpose, limitations, and potential risks |

| | - Consumer rights related to the use of AI system, including the Notice of data collection and the Right to opt-out |
|---|---|
| Clients (Law enforcement) | - User manuals for AI system<br>- Training materials for using the AI system<br>- Policies and procedures related to the use of the AI system<br>- Reports on the performance of the AI system, including any errors |
| Owner/Investors | - Financial statements, business plans, and pitch decks during the planning and development phases<br>-User growth and revenue during the testing and deployment phases<br>-Risk assessment reports and evaluation reports throughout the development and maintenance of the system. |
| Regulators | -Compliance reports, audits<br>-System risk assessment report<br>-Information on how the AI system was developed and tested<br>-Reports on the performance of the AI system<br>-Information on any ongoing monitoring and evaluation processes and other regulatory filings during the testing and deployment phases |
| Employees | -Employment contracts, employee handbooks, and job descriptions<br>-AI system specifications and technical documentation<br>-Policies and procedures related to the use and development of the AI system<br>-Training material on identifying and mitigating potential biases during the planning and development phases |
| Community | -Environmental impact reports and community outreach reports during the testing and deployment phases<br>-Information on how the AI system is being used by law enforcement and its potential impact on civil liberties<br>-Explanation of how the facial recognition system works, its purpose, limitations, and potential risks<br>-Opportunities for public input and feedback |

In the surveillance industry, stakeholders range widely, each with distinct interests. General surveillance stakeholders include end users or data subjects, who are often the public or individuals under surveillance, requiring transparency and assurance of their privacy rights. Clients, such as government agencies, private businesses, or law enforcement, depend on surveillance technology for security and operations, needing reliable and effective tools. Owners and investors in surveillance firms seek profitability and sustainable growth, focusing on financial returns and strategic business outcomes. Regulators play a critical role, enforcing legal and ethical standards to ensure surveillance practices comply with laws and respect civil liberties. Employees in these companies are responsible for the development, deployment, and maintenance of surveillance technologies, necessitating proper training and ethical guidelines. The broader community, affected by surveillance practices, looks for assurances about the impact on civil liberties, privacy, and societal norms. While these stakeholders are similar to those in companies like Clearview AI, the emphasis and specific concerns can vary based on the nature of the surveillance technology and its application. Each group's distinct interests and concerns shape the engagement and communication strategies of surveillance companies.

## 2.5 Continuous Monitoring and Improvement

Surveillance companies must have a continuous monitoring and improvement process in place to ensure their AI system remains effective and compliant with the mitigation plans and policies post-deployment. For a specific organization like Clearview AI, this would involve establishing a detailed plan for incident response, recovery, and change management, in line with MANAGE 4.1 guidelines. Risk assessment is another key component, where Failure Modes and Effects Analysis (FMEA) [28] could be employed to preemptively identify and address potential issues. The monitoring focus includes system performance, user interactions, error rates, and compliance with ethical standards. This provides insights into real-world usage, identifies improvement areas, and ensures legal and ethical alignment. Periodic evaluations, conducted using tools like AI Fairness 360, SHAP, or LIME [30, 31] can help assess fairness and explainability. These processes help in enhancing reliability, improving user experience, and maintaining ethical integrity. To ensure objectivity, parts of this process, particularly those involving ethical compliance and adherence to privacy laws, should be overseen by independent entities. This external oversight is key to an unbiased assessment and bolsters public confidence in the system.

Additionally, integrating stakeholder feedback and making data-driven adjustments will continually enhance the system's performance (MANAGE 4.2, 4.3). Finally, periodic reviews of the risk management process (GOVERN 1.4, 1.5), and secure decommissioning methods (GOVERN 1.6, 1.7) through methodology similar to DBAN [32] should be implemented. This coupled with event management systems like IBM QRadar [33] to monitor for any unusual activities post-decommissioning, would help maintain the system's integrity throughout its life cycle.

## 2.6   Benchmarking and Collaboration

A premise of this whole section has been that there is an accurate understanding of the performance of the system, but such knowledge requires regular benchmarking of both the AI system and the overall AI risk management practices [34, 35]. The AI RMF MEASURE 1.1, 2.3, 2.5, 2.6, and 2.11 functions, among others, provide clear guidance about what would need to be measured and monitored. The benchmarks outlined in Table 3 are vital for any company in the surveillance industry, including Clearview AI, to ensure their technology meets the high standards required for accuracy, fairness, and ethical deployment of facial recognition systems [17, 36, 37].

Table 3: Recommended Industry-Standard Benchmark Tests for Surveillance Technology Companies

| Benchmark Test | Description | Pass Criteria |
|---|---|---|
| NIST FRVT | Measures accuracy, speed, storage, and memory consumption | Evaluates for accuracy, processing power, and resource usage |
| IJB-C | Evaluates on challenging datasets, diverse poses, image quality | Tests adaptability in complex, real-world conditions. |
| MegaFace Challenge | Evaluates recognition with many distractor faces | Assesses accuracy in populated environments. |
| FERET | Measures performance across various facial expressions and conditions | Evaluates versatility and reliability across varied expressions and conditions. |
| DHS Biometric Rally | Focuses on operational use in high-throughput environments | Tests efficiency and accuracy in high-volume scenarios. |
| VGGFace2 | Evaluates algorithms on a diverse dataset | Ensures effectiveness and fairness across diverse demographics. |

# 3  Data as a Mechanism To Reduce Risk

Our application of the NIST AI RMF to surveillance technology has provided a structure for the many actions that surveillance companies should take. At the same time, it has highlighted some gaps in the current AI RMF. First, while the AI RMF is intended to be used across sectors, there are some potential issues when applying it to military, defense, or security technologies. For example, "risk of physical harm" could actually be a benefit for a system that is intended for use in conflict settings. Surveillance systems are increasingly used in wartime or conflict operations; surveillance companies play a pivotal role in modern warfare, as seen in Ukraine, where its software enables the identification of Russian military personnel and assists in locating missing individuals. However, the current version of the framework provides little guidance in addressing the unique requirements and risks of wartime situations. At the very least, it would be helpful to have supplementary guidance (e.g., about the relevance of the Law of Armed Conflict to AI systems) for AI in these sectors. More ambitiously, an AI RMF profile could provide relatively standardized guidance for these types of AI systems.

Second, the AI RMF arguably has an insufficient focus on data integration, management, storage, and deletion. Surveillance companies utilize data, including biometric data, from various sources to generate outputs, often dealing with sensitive information with substantial potential for misuse. The existing framework's guidance regarding the handling and merging of different types of personal data lacks the necessary level of detail, potentially exposing privacy vulnerabilities due to data aggregation (particularly for biometric data). To enhance the framework's effectiveness, it is crucial to conduct a thorough examination of various data integration scenarios and their potential consequences. To begin, the framework should establish specific rules for the combination of different types of data, particularly in cases involving sensitive information. We recommend that the data should be analyzed along the following categories:

a) Privacy Risk
b) Data Sensitivity
c) Regulatory Compliance
d) Integration Context
e) Potential Outcomes/Goals

Each data type carries its unique risks and potential for misuse, and the framework should provide guidelines on how these can be integrated safely and ethically. For instance, combining location data with purchasing history or internet browsing patterns can lead to intrusive profiling, which the RMF should address explicitly.

Relatedly, the AI RMF must place a greater emphasis on the temporal dimensions of data integration. Rather than restricting or outright banning data integration, one could instead restrict when and for how long such data combinations are permissible. This is especially crucial in sensitive areas like law enforcement or health diagnostics, where extended periods of data integration could lead to the erosion of individual privacy. To address this, the RMF should establish guidelines that promote short-term data integration in these sensitive scenarios, thereby mitigating long-term privacy risks. To ensure compliance with these guidelines, a robust audit trail is key. By recording each data transaction or integration event on a blockchain, we create an immutable and unalterable record that potentially ensures a secure, tamper-proof record.

Third, the RMF should offer a comprehensive approach to the retention and deletion of personal and contextual data. It has become cliché to talk about data as the "lifeblood" of an AI system, but it is nonetheless correct that the data themselves are a critical component of an AI system that must be assessed for risks, harms, and benefits. The retention policies need to be not only data-specific but also context-specific. For instance, biometric data used in a criminal investigation might have a different retention period compared to biometric data used in a healthcare setting. This distinction is vital because the risks and necessities vary greatly between these contexts. The framework should guide organizations on how to evaluate and determine these retention periods based on a combination of ethical considerations, legal requirements, and practical needs. In addition to focusing on the retention of data, the RMF should also highlight the importance of data deletion. It should include guidelines for context-aware deletion strategies, where data associated with specific events or purposes, such as a criminal investigation or a clinical study, is automatically designated for deletion post the conclusion of these events. This ensures that data isn't retained unnecessarily, reducing the chances of privacy violations and potential misuse.

## 3.1 Impact of Parent-Subsidiary Relationships on AI Governance

The NIST AI RMF provides a comprehensive set of guidelines for managing risks associated with AI systems. However, it falls short in addressing the intricacies involved in parent and subsidiary company structures, particularly within surveillance companies that handle Personally Identifiable Information (PII). The AI RMF focuses on technology that is designed, developed, and deployed in a single integrated process, and so the distributed nature of present-day AI development can lead to gaps or problems that are not necessarily identified in the current AI RMF [38]. This oversight becomes critically evident when examining the legal and ethical challenges of data transfer and governance in these corporate structures. For example, Clearview AI includes foreign investors and thus introduces factors such as cross-border data flow and the impact of foreign ownership on data security and privacy. More generally, a significant concern arises from the potential disconnect between the terms under which PII is collected by a subsidiary and the subsequent use of this data by a parent company. Unless contracts are carefully written (and with the data subjects in mind, rather than corporate profits), the parent company can have surprising and intrusive levels of access to, or control over, the subsidiary's PII. This oversight can lead to scenarios where the PII is used or managed in ways that the data subjects did not consent to, raising serious questions about privacy and user rights.

To address these challenges, the NIST AI RMF should incorporate provisions that ensure any entity acquiring PII - whether through contract, merger, acquisition, or corporate structure - is bound by the original terms of data collection. Assuming that the data were collected ethically (e.g., actual informed consent), then this restriction would create a continuous chain of consent and responsibility, ensuring that data governance standards are upheld despite organizational changes. Furthermore, the framework should explicitly outline the responsibilities of parent companies in managing PII. This includes enforcing stringent data storage, retention, and deletion practices, even if the parent company does not actively use the data. Such measures are crucial for maintaining the integrity and confidentiality of PII and for aligning with evolving privacy regulations.

The absence of these specific guidelines in the NIST AI RMF presents a notable gap, especially given the increasing prevalence of data-driven business models and the complex corporate structures in the tech industry. By updating the RMF to address these issues, the NIST can provide clearer direction for companies in managing PII ethically and legally, particularly in situations where data governance responsibilities are transferred as a result of corporate restructuring.

# 4 Conclusions

Utilizing the NIST AI Risk Management Framework (RMF) unveils significant insights into facial recognition surveillance technology, highlighting aspects that might otherwise remain undiscovered. The approach not only enhances our understanding of the technology's accuracy but also comprehensively examines its societal, cultural, and systemic impacts. The RMF, with its standardized application across various fields, provides an opportunity for surveillance companies like Clearview AI to contribute to this evolving standard, ensuring a more responsible AI ecosystem. The benefits of employing RMF extend beyond improving the accuracy of facial recognition systems; they also include the development of ethical business practices and the provision of avenues for recourse.

To enhance AI risk management, we introduced the AI System Risk Categorization Matrix (ASRCM) for streamlined risk evaluation and prioritization, coupled with a robust post-deployment monitoring plan for ongoing system refinement and adaptability. These steps, informed by the RMF, ensure substantial improvements in the organization's operations, leading to a safer, more dynamic, and ethically responsible AI environment. That being said, we acknowledge that risk management is a continual process, and so we conclude by presenting some additional strategic steps that a company could consider to maintain a secure, effective, and responsible AI ecosystem:

- Emerging Technologies and Standards: Clearview AI should actively monitor the latest developments in AI technology, emerging standards, regulations, and best practices. Investing in research and development will help the company ensure that its AI systems remain innovative, efficient, and in line with industry trends [39].
- Establish an Ethical Advisory Board: The organization could benefit from creating an ethical advisory board composed of experts from diverse fields, such as technology, law, ethics, and social sciences. This board would provide guidance on ethical considerations, ensuring that the company's AI-driven surveillance technologies align with societal values and expectations [40, 41].
- Promote Environmental Sustainability: Clearview AI should evaluate the environmental impact of its AI-driven technologies and strive to minimize energy consumption and carbon footprint throughout development, deployment, and maintenance. Adopting sustainable practices will allow the company to contribute to global efforts against climate change while promoting responsible AI use [42, 43].

By incorporating these strategic steps into their long-term planning, Clearview AI can strengthen its AI risk management practices. Future considerations should include governance and policy in data management, especially in sensitive areas like wartime operations.

**ACKNOWLEDGMENTS**

On behalf of all authors, the corresponding author states that there is no conflict of interest. The authors have no relevant financial or non-financial interests to disclose.